# Sub-Hz Linewidth Photonic-Integrated Brillouin Laser

Sarat Gundavarapu[1], Ryan Behunin[2], Grant M. Brodnik[1], Debapam Bose[1], Taran Huffman[1], Peter T. Rakich[3], and Daniel J. Blumenthal[1†]

[1] Department of Electrical and Computer Engineering, University of California at Santa Barbara, Santa Barbara, CA 93106 USA
[2] Department of Physics and Astronomy, Northern Arizona University, Flagstaff, Arizona 86011 USA
[3] Department of Applied Physics, Yale University, New Haven, CT 06520 USA

† Corresponding author (danb@ucsb.edu)

(Date 10/xx/2017)

## ABSTRACT

Photonic systems and technologies traditionally relegated to table-top experiments are poised to make the leap from the laboratory to real-world applications through integration, leading to a dramatic decrease in size, weight, power, and cost[1]. In particular, photonic integrated ultra-narrow linewidth lasers are a critical component for applications including coherent communications[2], metrology[3–5], microwave photonics[6], spectroscopy[7], and optical synthesizers[1]. Stimulated Brillouin scattering (SBS) lasers, through their unique linewidth narrowing properties[8], are an ideal candidate to create highly-coherent waveguide integrated sources. In particular, cascaded-order Brillouin lasers show promise for multi-line emission[14], low-noise microwave generation[6] and other optical comb applications. To date, compact, very-low linewidth SBS lasers have been demonstrated using discrete, tapered-fiber coupled chip-scale silica[9,10] or $CaF_2$[11] microresonators. Photonic integration of these lasers can dramatically improve their stability to environmental and mechanical disturbances, simplify their packaging, and lower cost through wafer-scale photonics foundry processes. While single-order silicon[12] and cascade-order chalcogenide[13] waveguide SBS lasers have been demonstrated, these lasers produce modest emission linewidths of 10-100 kHz and are not compatible with wafer-scale photonics foundry processes. Here, we report the first demonstration of a sub-Hz (~0.7 Hz) fundamental linewidth photonic-integrated Brillouin cascaded-order laser, representing a significant advancement in the state-of-the-art in integrated waveguide SBS lasers. This laser is comprised of a bus-ring resonator fabricated using an ultra-low loss (< 0.5 dB/m) $Si_3N_4$ waveguide platform. To achieve a sub-Hz linewidth, we leverage a high-Q, large mode volume, single polarization mode resonator that produces photon generated acoustic waves without phonon guiding. This approach greatly relaxes phase matching conditions between polarization modes and optical and acoustic modes. By using a theory for cascaded-order Brillouin laser dynamics[14], we determine the fundamental emission linewidth of the first Stokes order by measuring the beat-note linewidth between and the relative powers of the first and third Stokes orders. Extension of these high performance lasers to the visible and near-IR wavebands is possible due to the low optical loss of silicon nitride waveguides from 405 nm to 2350 nm[15], paving the way to photonic-integrated sub-Hz lasers for visible-light applications including atomic clocks and precision spectroscopy.

# I. INTRODUCTION

Photonic-integrated sub-Hz linewidth lasers will enable miniaturization, lower cost and low power consumption, for a wide range of applications including coherent communications[2], spectroscopy[7], metrology[4], sensing[16,17], atomic clocks[3] and optical frequency comb generators[18]. Stimulated Brillouin scattering (SBS) lasers are capable of sub-Hz emission by greatly reducing the fundamental linewidth[19–22] and driving the pump phase diffusion component[8] well below this linewidth. While discrete component Brillouin lasers can produce highly-coherent sub-Hz emission, this performance has not been demonstrated in photonic-integrated SBS lasers, where improvements in stability, efficiency, and cost can be made.

To date, demonstrations of sub-Hz SBS lasing utilize discrete photonic components[19,11,23,24] and complex feedback control and stabilization mechanisms. These lasers are environmentally and vibration sensitive, consume high power, occupy large footprints and are costly to manufacture. Waveguide based photonic-integrated SBS lasers, with the performance of discrete systems, can address these challenges and allow integration of the SBS laser into higher-function photonic circuits. However, the path to integration has remained elusive. The most compact low linewidth designs to date, utilize non-integrated tapered-fiber coupled silica micro-disk resonators[9]. Single and cascaded-order waveguide integrated SBS lasers with kHz linewidths have been demonstrated using suspended silicon waveguides[12] and chalcogenide waveguides wafer-bonded to silicon[13,25] using non-wafer-scale foundry compatible processes. However, existing and emerging applications that require sub-Hz linewidths, will benefit from the mechanical and environmental stability brought about by photonic integration, and the reduction of low frequency technical noise sources that are inherent in non-integrated chip-scale solutions. Foundry-scale integration process will also provide these applications the benefits of lower cost and integration of sub-Hz lasers with systems-on-chip.

In this paper we report the first demonstration of a sub-Hz (~ 0.7 Hz) fundamental linewidth photonic-integrated waveguide laser. The laser is a cascaded-order SBS design[26] fabricated in a $Si_3N_4$ waveguide integration platform. We leverage a high-Q, large mode volume, single polarization mode resonator that produces photon generated acoustic waves without phonon guiding. This approach greatly relaxes phase matching conditions between polarization modes and optical and acoustic modes easing dispersion engineering challenges present in other designs[9,12,27,]. Rapidly decaying unguided phonons suppress the contribution of pump phase diffusion component to the linewidth, resulting in a laser performance determined by the fundamental sources of noise. The fundamental linewidth is reduced below 1 Hz using a long (74 mm), high loaded Q (>28 Million) resonator capable of supporting high intra-cavity intensities. We determine sub-Hz linewidth laser emission using a theory for cascaded-order Brillouin laser dynamics[14] that only requires measuring the first and third Stokes order beat-note linewidth the relative measured Stokes powers. Using this method, we find a sub-Hz first Stokes order linewidth of ~ 0.7 Hz. We also discuss how this method can be used to quantify changes in the resonator parameters change when moving from below and above lasing threshold by comparing the difference between the predicted and measured linewidths. This provides a valuable technique for determining how parameters like loaded Q and bus to ring coupling coefficient change from the cold cavity to the hot cavity state and can lead to improved operating designs.

Due to the low loss of $Si_3N_4$ waveguides across an extremely broad range of wavelengths (405 nm – 2350 nm)[15,28,29] this level of performance has the potential to translate to the visible and other wavebands, opening the potential for compact sub-Hz source to a wealth of applications.

This paper is structured as follows. In Section II, we present the theoretical expression for the linewidth[14] of a cascaded-order Brillouin laser Stokes tone. We present an overview of the properties our waveguide Brillouin laser and describes how these properties enable cascaded Brillouin lasing and narrow linewidth emission. The measurements and analysis of the inter-Stokes order beat note and individual Stokes linewidths is reported in Section III. In Section IV, we discuss the possible applications and future directions of our work.

## II. THEORY OF THE FUNDAMENTAL LINEWIDTH OF A BRILLOUIN LASER

The fundamental linewidth of a cascaded-order Brillouin laser is determined by laser resonator quality, the thermal and quantum fluctuations of the optical mechanical fields participating in Brillouin scattering, and the coherent occupation of all the lasing orders. For the $m^{th}$ Stokes order in a cascaded-order Brillouin laser, this fundamental linewidth $\Delta \nu_m$ is given by[17]

$$\Delta \nu_m = \frac{1}{4\pi N_m} \left[ \gamma_m \left( N_m^{th} + n_{m-1}^{th} + 1 \right) + 2\mu_m N_{m+1} \left( n_m^{th} + n_{m-1}^{th} + 1 \right) \right]. \quad (1)$$

Here, $\gamma_m$, $\mu_m$, $N_m$ and $N_m^{th}$ are the respective decay rate, Brillouin amplification rate per pump photon, and coherent intracavity and thermal photon numbers, all defined for the $m^{th}$ Stokes order. The thermal occupation number, $n_m^{th}$, is defined for the phonon mode mediating Brillouin scattering between the $(m-1)^{th}$ and $m^{th}$ orders. The parameter $\mu_m$ can be expressed in terms of more familiar quantities with the following relation

$$\mu_m = \frac{\hbar \omega_m c^2}{2 n_g^2 L} \left( \frac{g_{Bm}}{A_{eff}} \right) \quad (2)$$

where $\omega_m$ and $g_{Bm}$ are the respective angular frequency and bulk Brillouin gain coefficient of the $m^{th}$ mode. Other parameters are the group index ($n_g$), the group index, the resonator length $(L)$, the effective opto-acoustic overlap area ($A_{eff}$), and $c$ and $\hbar$, the speed of light and the reduced Planck constant respectively. In addition, Equation 1 can be cast in terms of more readily measurable quantities by using the relation $\gamma_m/N_m = \hbar \omega_m^3/(Q_T Q_E P_{Sm})$, where $Q_T$ and $Q_E$ are the loaded and external resonator quality factors and $P_{Sm}$ is the emitted power from the $m^{th}$ laser order.

The fundamental linewidth for the $m^{th}$ Stokes order in Eq. (1) can be reduced using a high-Q (small $\gamma_m$), large mode volume ($LA_{eff}$), resonator that supports high intra-cavity optical intensities ($N_m \gg 1$ and $N_m^{th} \ll 1$). For a Brillouin laser, the fundamental linewidth is dominated by the thermomechanical noise arising from the thermal phonon number $n_m^{th}$ (~570 at room temperature). For our system with relatively large gain bandwidth, the transferred pump noise is compressed by a factor of $(1+ \Gamma/\gamma_m)^2$, where $\Gamma$ is the gain bandwidth (or decay rate of the acoustic mode), ensuring that the linewidth component that dominates in other waveguide designs, is negligible compared to the fundamental linewidth in Eq. (1).

# LASER OVERVIEW

We implement a Brillouin laser from an integrated waveguide ring-bus resonator of the type shown in Fig. 1(a) and described in Ref. 26. To drive laser emission, pump light, resonant with an optical mode of the laser cavity, is injected into the waveguide bus (see Fig. 1 (a)). High intra-cavity pump intensities, generated by resonant build up, lead to spontaneous Stokes emission. This spontaneous Stokes emission populates the S1 optical mode that lases when the round-trip cavity losses are compensated by the SBS amplification. Cascaded-order lasing, with laser emission at several distinct wavelengths, can occur as this process repeats when the pump power is further increased, and the first Stokes pumps the second Stokes mode, the second pumps the third, and continues, as illustrated in Fig. 1(b).

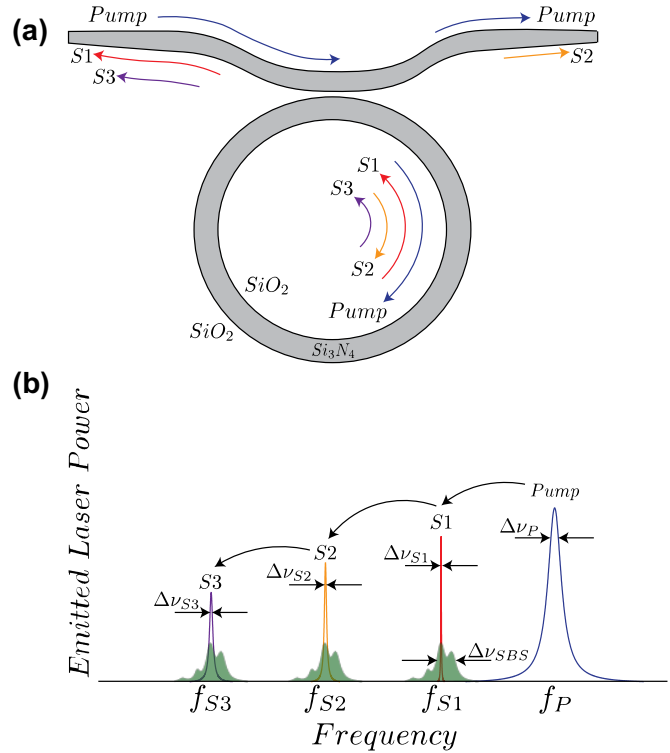

Fig. 1. *Narrow linewidth cascaded-order $Si_3N_4$ Brillouin laser* (a) Ring-bus Brillouin laser resonator schematic. Pump light (blue) injected into the bus waveguide drives cascaded Brillouin lasing. Odd orders emit from the reflection port, while even orders emit through the transmission port (b) Energy transfer illustration of cascaded Brillouin lasing and cascaded order linewidth. Higher emission powers produce greater line narrowing as described by Eq. (1). The relatively broad pump linewidth is highly compressed in laser emission. Brillouin gain spectrum is shown (green curve).

The unique properties of our laser resonator facilitate highly-coherent emission and allow lasing over a wide range of pump wavelengths. These properties include ultra-low waveguide losses, large resonator mode volume, and a broadened Brillouin gain bandwidth. In particular, we achieve gain broadening by harnessing unguided (essentially bulk) phonons to mediate energy transfer from the pump to cascaded laser orders. Equation (1) shows how these properties can enhance laser coherence (lower linewidth). A large mode volume ($LA_{eff} \sim 2.07 \times 10^{-12}$ m$^3$) and broadened gain bandwidth, foster large intra-cavity photon number ($N_m \gg 1$) and lower the Brillouin amplification rate ($\mu_m$). The long resonator length ($L$ = 74 mm), made possible by the Si$_3$N$_4$ waveguide platform, yields a 2.72 GHz free spectral range (FSR), four times smaller than the Brillouin frequency shift $\Omega_B/2\pi$ =10.93 GHz. This FSR, combined with the relatively large gain bandwidth generated by unguided phonons of 200 MHz (green curve in Fig. 1(b)) is a factor four times larger than that of other reported designs[11,14] and guarantees that the input pump is always within ~1.2 GHz of a lasing resonance.

## III. DETERMINATION OF THE FUNDAMENTAL LINEWIDTH

We determine the fundamental laser linewidth using the experimental setup shown in Fig. 2(a). The beat note phase noise power spectrum $L(\nu)$, between the first and third Stokes tones S1 and S3, determined by fundamental sources of noise is given by[30]

$$L(\nu) = \frac{\Delta \nu_{S1-S3}}{2\pi \nu^2}, \quad (3)$$

where $\Delta \nu_{S1-S3}$ is the linewidth of the S1, S3 photo-mixed beat note on a detector and $\nu$ is the frequency offset from this beat note.

The noise dynamics of cascaded Brillouin lasers[14] show that the fundamental linewidth $\Delta \nu_{S1-S3}$ can be expressed as the sum of the individual linewidths of the participating Stokes tones[14], $\Delta \nu_{S1}$ and $\Delta \nu_{S3}$, as $\Delta \nu_{S1-S3} = \Delta \nu_{S1} + \Delta \nu_{S3}$, and additionally yields the dependence of these linewidths on the emitted powers $P_{S1}$ and $P_{S3}$. For third-order cascading, the individual Stokes order linewidths are directly related to the measured value for $\Delta \nu_{S1-S3}$ and the relative optical powers through the following relations[14]

$$\Delta \nu_{S3} = \frac{\Delta \nu_{S1-S3}}{\left(1 + \frac{3 P_{S3}}{P_{S1}}\right)} \text{ and } \Delta \nu_{S1} = \frac{\Delta \nu_{S1-S3}}{\left(1 + \frac{P_{S1}}{3 P_{S3}}\right)}. \quad (4)$$

These relations, between the measured linewidth of beat note and the linewidths of individual Stokes orders, depend only on the ratio of the individual optical powers and not the absolute optical powers of the Stokes orders. This important property removes the requirement to know the cavity parameters under lasing conditions and minimizes measurement errors of the on-chip Stokes order powers.

We optically pump the laser with a fiber coupled tunable CW source. For an on-chip pump power of 125 mW, we observe cascaded Brillouin lasing to three Stokes orders as shown in Fig. 3(a). The first and third Stokes orders are measured using a circulator at the resonator reflection port and separated from residual pump power using a fiber Bragg grating (FBG) optical filter. To suppress sources of technical noise, the pump laser is locked to a laser cavity resonance using a Pound-Drever-Hall control loop[31]. We measure the phase noise power spectrum of the 21.8 GHz beat note using photo-mixed first and third order Stokes tones on a photodetector, amplified using a low noise RF amplifier, using a signal source analyzer (Keysight Model E5052B), and a microwave downconverter (Keysight Model E5053A) for offset frequencies $\nu$ (100 Hz - 40 MHz).

We find a 2.76 Hz upper bound on the fundamental beat note linewidth $\Delta \nu_{S1-S3}$, obtained by finding the smallest value for $\Delta \nu_{S1-S3}$ such that Eq. 3 intersects the phase noise power spectrum at a single point (see black dotted line in Fig. 3b). Put into words, we assume that the technical and fundamental sources of noise are uncorrelated, and therefore these sources of noise add in quadrature. Consequently, if technical sources of noise are completely eliminated, the fundamental noise can be no larger than fundamental noise described by Eq. 3 with $\Delta\nu_{S1-S3}$ = 2.76 Hz. We calculate the fundamental linewidths of S1 and S3 using the ratio of the respective measured powers, 8 mW and 0.9 mW coupled to a fiber, and the beat note linewidth $\Delta\nu_{S1-S3}$ = 2.76 Hz, as shown in the green and red dashed lines in Fig. 3(b). Equation (4) yields the first and third Stokes orders linewidths to be $\Delta\nu_{S1}$ = 0.7 Hz and $\Delta\nu_{S3}$ = 2.06 Hz, demonstrating sub-Hz emission for the first Stokes order.

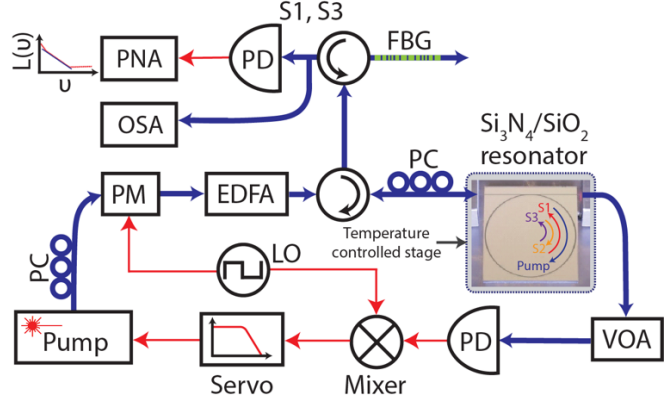

Fig. 2. ***Brillouin laser linewidth characterization setup.*** *Experimental setup of PDH locked Si$_3$N$_4$ based Brillouin laser: PC – Polarization controller, PM – Phase modulator, EDFA – Erbium doped fiber amplifier, PD – Photodiode, FBG – Fiber Bragg grating to filter reflected pump, LO – Local oscillator (40 MHz), VOA – Variable optical attenuator, PNA – Phase noise analyzer (Keysight E5052B signal source analyzer with E5053A down converter); OSA – optical spectrum analyzer.*

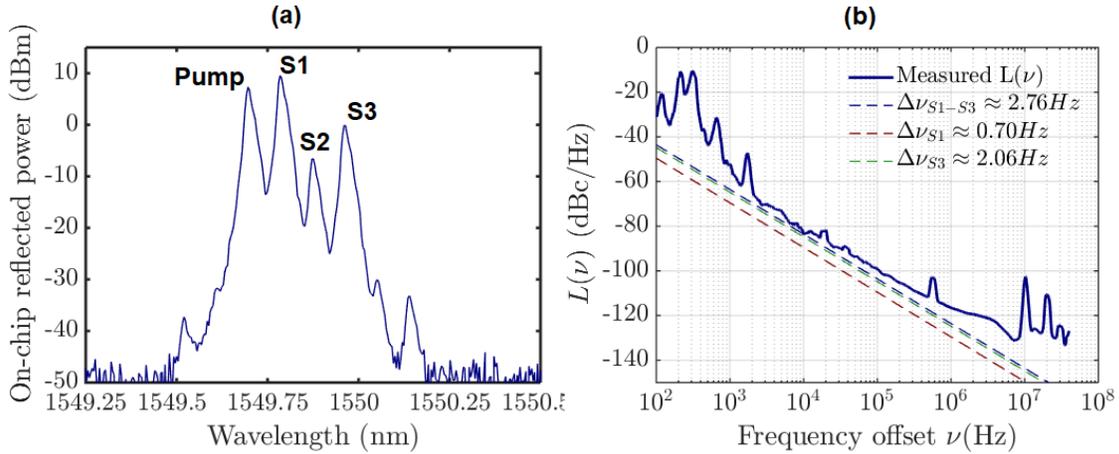

Fig. 3. ***Phase noise characterization and fundamental linewidth of Brillouin laser.*** *(a) Measured reflected powers of pump and first three Stokes orders from the resonator (b) Phase noise of beat note between first and third Stokes components indicating a 0.70 Hz fundamental linewidth of S1.*

An important set of parameters that determine the emission linewidths are related to the change in resonator properties as the cavity transitions from a cold-state below threshold to a hot-state at the operating point above threshold. Changes in the cavity physical parameters occur due to temperature dependent waveguide index changes and thermal expansion that impact the waveguide group index, the ring resonator mode volume and the bus to ring directional coupler coupling parameters. In particular, the ring-bus coupling coefficient is exponentially sensitive to the coupler gap[32], and a small degree of thermal expansion can significantly change the respective loaded and external resonator quality factors $Q_T$ and $Q_E$. Additionally, accurate measurement of on-chip cavity powers, using off chip measurements, are susceptible to error in calibrated fiber-chip coupling loss, reflections at chip facets, wavelength filtering and other measurement uncertainties. Therefore, linewidth calculations using Eq. (1) can lead to significant errors under lasing conditions, when using cold cavity parameters and off-chip absolute optical power measurements.

There are two main benefits in using Eq. (4) in determining the Stokes orders linewidths. First, the linewidth calculation only relies on the measured beat note phase noise spectrum and the relative Stokes order powers, providing a measurement technique that normalizes out the loaded and external cavity Q factors, and Stokes optical powers. This leads to an invaluable tool for determining the changes in coupler and ring parameters in the lasing condition.

An example is given in Table 1. Measurement of the S1, S3 beat note linewidth at 2.76 Hz, and the ratio of the off-chip Stokes 1 and Stokes 3 powers at 8.0 mW/0.9 mW, yields through Eq. (4) linewidths of 0.7 Hz and 2.06 Hz respectively. While utilizing Eq. (1) and the measured cold cavity loaded and external Q factors[26], and off-chip Stokes 1 and Stokes 3 powers of 8.0 mW and 0.9 mW, yields linewidth estimates of 1.78 Hz and 5.28 Hz respectively.

*Table 1: Summary of $Si_3N_4$ laser linewidth measurements and properties*

| T (K) | λ (nm) | $Q_T/Q_E$ (million) | $P_{S1}/P_{S3}$ (mW) | $\Delta v_{S1-S3}$ (Hz) using $Eq.$ (3) | $\Delta v_{S1}/\Delta v_{S3}$ (Hz) using $Eq.$ (4) | $\Delta v_{S1}/\Delta v_{S3}$ (Hz) using $Eq.$ (1) |
|---|---|---|---|---|---|---|
| 300 | 1550.0 | 28.5/63.5 | 8.0/0.9 | 2.76 | 0.70/2.06 | 1.78/5.28 |

The results in Table 1 can be understood as follows. Linewidths calculated using Eq. (4) require knowledge of only the beat note linewidth and the ratio of the optical powers, measured off-chip for S1 and S3. The resonator cavity details, including the volume, group index, and ring-bus power coupling factor do not need to be known, assuming they are constant over the frequency separation between the S1 and S3 Stokes waves. As a result, errors in calibrating the measured off-chip power to the on-chip power are normalized out using the power ratio, assuming errors are the same for both Stokes frequencies. Alternatively, using Eq. (1) to predict the linewidths, requires accurate knowledge of the mode volume, the ring-bus power coupling coefficient and the absolute on-chip Stokes powers. These values are difficult to calibrate, and will change as a function of thermal heating of the cavity as it moves from a cold-cavity state to hot-cavity state when the system is lasing and intra-cavity photon density builds up significantly. Additionally, the calibration uncertainties in measuring off-chip powers accurately, and power loss due to reflections, adds additional linewidth errors in using Eq, (1). However, once the Stokes order linewidths have been determined using Eq. (4), the hot-cavity resonator parameters can then be calculated using these linewidth results in combination with Eq. (1). Table 1 indicates the errors when using cold-cavity measurements and estimates. With measured off-chip Stokes powers of 8.0 mW and 0.9 mW for S1 and S3, and a beat note linewidth of 2.76 Hz from phase noise measurements, the Stokes order linewidths are calculated at 0.7 for S1 and 2.06 for S3. Conversely, if the loaded and unloaded cold-cavity Qs of 28.5 Million and 63.5 Million are used with these same off-chip measured powers, the estimate of Eq. (1) yields an S1 linewidth of 1.78 Hz and S3 linewidth of 5.28 Hz, a factor of 2.5 increase, with the same ratio (down to two decimal places). This discrepancy is due to the change in cavity parameters between cold and hot operation as described above, and is significant. Using the linewidths calculated using Eq. (4), we find from Eq. (1), a hot-to-cold power coupling ratio of 0.39 for both the first and third Stokes orders suggesting a 39% reduction of the ring-bus coupling in the hot cavity. The combination of exponential temperature sensitivity of the coupler gap[32] and thermal changes in the cavity mode volume helps explain these observations. This result leads to a new design tool that will enable coupler and cavity design corrections to be made to accommodate for hot-cavity lasing operating conditions.

## IV.    DISCUSSION

In this paper, we report the first waveguide-integrated sub-Hz Brillouin laser. Comprised of an integrated ring-bus $Si_3N_4$ waveguide resonator, this laser combines low-optical losses, a large mode volume, and a broad Brillouin gain bandwidth to produce highly-coherent cascaded-order laser emission. The long resonator length and single polarization operation, made possible by this unique waveguide platform, as well as the relatively broad gain bandwidth, minimizes the need for dispersion engineering and enables high intra-cavity intensities. Consequently, narrow-linewidth Brillouin lasing can be achieved over a broad range of pump wavelengths. By combining heterodyne measurements of distinct Stokes order beat notes, measurements of the emitted laser power, and a new theory[14] describing the noise properties of cascaded-order Brillouin lasers, we determined the fundamental laser linewidth. Application of this theory allows calculation of the Stokes order linewidths using the ratio of Stokes powers only, eliminating errors in calibrating measurement of on-chip power, and taking into account linewidth noise dynamics that feedback from higher order lasing Stokes orders. This analysis, yields a 0.7 Hz linewidth for the first Stokes order.

Using cascaded Brillouin laser theory, measurements of the emitted powers and heterodyne beat notes can be used to estimate hot-cavity properties (see Tab. I). We estimate a 39% reduction in the ring-bus power coupling coefficient, when the laser operation moves from below to its operating point above threshold, implying that the resonator quality factor improves with high-intracavity powers. Given the exponential sensitivity of the external coupling to the directional coupler gap, this hypothesis gives a plausible explanation for our data, where a small degree of thermal expansion may have a dramatic effect of the resonator Q factor.

The foundry-compatible platform and fabrication of this laser make it possible to create high-performance light sources with reduced size, cost, and power consumption. This compatibility allows this chip-scale laser to be integrated with a wide variety of $Si_3N_4$ and silicon photonic waveguide active[33] and passive[34,35] components. Further integration with heterogeneous silicon/III-V tunable lasers[21,36] can leverage the property of pump phase noise and RIN reduction in Brilloiun lasers[31], to further pave the way to low cost, tunable highly coherent Brillouin lasers. Combined with a tunable pump, wavelength locked to the laser resonator, a compact, low-cost highly coherent source for ITU grid and gridless[37] coherent terabit communications systems is possible. This Brillouin laser is also an ideal source for microresonator based Kerr frequency combs[38], as the technical noise of the comb is limited by noise inherent in the pump laser, with application to WDM compact sources for coherent terabit communications[5]. Moreover, the large transparency window, from 405 nm - 2350 nm, of the $Si_3N_4$ platform paves the way for integrated sub-Hz lasers that can extend to the visible wavelength range. Operation in the visible and near-IR spectrum promises compact coherent sources for spectroscopy[39] and atomic clocks[3]. The $Si_3N_4$ integration platform is readily compatible with previously demonstrated chip-scale photonic components[33,34,40] leading to higher complexity circuits that leverage sub-Hz performance, and wafer-scale foundry CMOS compatible processes.

## ACKNOWLEDGEMENTS

This material is based upon work supported by the Defense Advanced Research Projects Agency (DARPA) and Space and Naval Warfare Systems Center Pacific (SSC Pacific) under Contract No. N66001-16-C-4017. The views and conclusions contained in this document are those of the authors and should not be interpreted as representing official policies of DARPA or the U.S. Government. We would like to thank Karl Nelson, Matthew Puckett, and Jianfeng Wu for extremely helpful discussions. We also thank Paul Morton from Morton Photonics for the use of the Morton Photonics laser as a pump, Archita Hati and Craig Nelson from NIST for their helpful discussions and help in measuring phase noise at the NIST laboratories and Catia Pinho at the Instituto de Telecomunicações (IT) for help with cavity Q characterization. We also thank Biljana Stamenic for help in processing samples in the UCSB nanofabrication facility, and at Honeywell, William Renninger for help with the measurement techniques for Brillouin gain profiles and Joe Sexton, Jim Hunter and Dane Larson for the cladding deposition, pre-cladding preparation and anneal process.